%% file: MGMR3D-v8.tex
\documentclass[pdftex,dvipsnames,usenames,aps,superscriptaddress,twocolumn,amsmath,showkeys]{revtex4-1}

\usepackage{hyperref}
\usepackage{float}
\usepackage{dcolumn}
\usepackage{graphicx}
\usepackage{color}
\usepackage{amssymb}
\usepackage{bold-extra}
\usepackage{caption}
\usepackage{subcaption}
\usepackage{verbatim}

\input{Preambule-Refs}

\newcolumntype{d}[1]{D{.}{.}{#1} }
\newcolumntype{h}[1]{D{-}{-}{#1} }

\def\vB{{\bf{v}\times\bf{B}}}
\def\vvB{{\bf{v}\times\left(\bf{v}\times\bf{B}\right)}}

\def\beq{\begin{equation}}
\def\eeq{\end{equation} }
\def\bea{\begin{eqnarray}}
\def\eea{\end{eqnarray}}

\def\appref#1{Appendix~\ref{app:#1}}
\def\applab#1{\label{app:#1}}  
\def\figref#1{Fig.~\ref{fig:#1}}
\def\figlab#1{\label{fig:#1}}  
\def\eqref#1{Eq.~(\ref{eq:#1})}

\def\eqlab#1{\label{eq:#1}}
\def\tabref#1{Table~\ref{tab:#1}}
\def\tablab#1{\label{tab:#1}}  
\newcommand*{\secref}[1]{Section~\ref{sec:#1}}
\newcommand*{\seclab}[1]{\label{sec:#1}}
\newcommand{\Omit}[1]{}

\def\Xmax{X_{\rm max}}
\def\Xrh{X_{\rm z}}
\def\Prog#1{} 
\def\SName{MGMR3D}

\def\KVI{University of Groningen, KVI Center for Advanced Radiation Technology, Groningen, The Netherlands}

\def\VUB{Vrije Universiteit Brussel, Dienst ELEM, Brussels, Belgium}

\begin{document}

\title{Analytic calculation of radio emission from parameterized extensive air showers, a tool to extract shower parameters}

\author{O.~Scholten}  \email[]{scholten@kvi.nl} \affiliation{\KVI}   \affiliation{\VUB}
\author{T.~N.~G.~Trinh} \email[]{t.n.g.trinh@rug.nl}  \affiliation{\KVI}
\author{K.D.~de Vries} \affiliation{\VUB}
\author{B.M.~Hare}  \affiliation{\KVI}

\date{\today}

\begin{abstract}
The radio intensity and polarization footprint of a cosmic-ray induced extensive air shower is determined by the time-dependent structure of the current distribution residing in the plasma cloud at the shower front.
In turn, the time dependence of the integrated charge-current distribution in the plasma cloud, the longitudinal shower structure, is determined by interesting physics which one would like to extract such as the location and multiplicity of the primary cosmic-ray collision or the values of electric fields in the atmosphere during thunderstorms.
To extract the structure of a shower from its footprint requires solving a complicated inverse problem.
For this purpose we have developed a code that semi-analytically calculates the radio footprint of an extensive air shower given an arbitrary longitudinal structure. This code can be used in a optimization procedure to extract the optimal longitudinal shower structure given a radio footprint.
On the basis of air-shower universality we propose a simple parametrization of the structure of the plasma cloud. This parametrization is based on the results of Monte-Carlo shower simulations.
Deriving the parametrization also teaches which aspects of the plasma cloud are important for understanding the features seen in the radio-emission footprint.
The calculated radio footprints are compared with microscopic CoREAS simulations.
\end{abstract}


\maketitle


\section{Introduction}

When a high-energy cosmic particle impinges on the atmosphere of Earth, it creates an extensive air shower (EAS). The electrons and positrons in the plasma cloud at the shower front are deflected in opposite directions due to the Lorentz force caused by the geomagnetic field. This creates an electric current. Since the number of particles in the EAS changes with penetration depth the electric current in the plasma cloud changes as a function of height in the atmosphere. This varying current emits radio waves~\cite{Sch08,Wer08,Wer12,Sch13} where the intensity pattern on the ground, the intensity footprint, depends on the variation of the current with height. The penetration depth where the particle number reaches its maximum, $\Xmax$, strongly depends on the specifics of the first collisions, in particular their multiplicity and thus the mass of the initiating cosmic ray. Different values of $\Xmax$ result in differences in the longitudinal variation of the currents which is reflected in the intensity footprint. Thus $\Xmax$ can be reconstructed on the basis of the footprint which allows for a determination of the mass composition of cosmic rays~\cite{Bui14} for fair-weather conditions.

In refs.~\cite{Sche15,Tri15}
a new method is introduced to determine the electric fields in the atmosphere during thunderstorm conditions by using the radio footprint from an EAS.
The basic principle is the same as used in determining $\Xmax$ for air showers recorded under fair-weather conditions (fair-weather events). During thunderstorm conditions an additional strong variation of the current with height is induced by the thunderstorm electric field~\cite{Sche15,Tri15}, which also varies with height in direction and magnitude. During such conditions the effect of the atmospheric electric field on the current can be dominant. While it is sufficient to consider only the intensity footprint for fair weather showers, the footprint for all Stokes parameters~\cite{Sche14,Scho16} are necessary for a complete mapping of the fields in the atmosphere~\cite{Tri16,Tri17}.

Only a single parameter, $\Xmax$, needs to be extracted from the radio-footprint for a fair-weather event. However, for a shower recorded under thunderstorm conditions (thunderstorm event) there are many more, order of 10, where the precise number of parameters depends on the level of sophistication of the modeling of the electric fields in the atmosphere. Therefore a simple grid search algorithm suffices to extract the value of $\Xmax$ for a fair-weather event, while such a grid search is totally impractical for a thunderstorm event.
To make such a parameter search more efficient one needs to be able to deterministically calculate the radio-footprint given the structure of the shower, such that an infinitesimal change in the shower parameters results in an infinitesimal change in the radio footprint.
In addition, it is convenient if a single calculation takes little computer resources, as such a calculation has to be iterated many times to find the optimum.
These two conditions are not met by
the presently available microscopic codes, CoREAS~\cite{Hue13} and ZHAireS~\cite{Alv12}. Since both of these codes are based on a Monte-Carlo calculation of the EAS, changing a single shower parameter will, in general, affect the complete shower evolution in a non-deterministic way. Furthermore, the codes are rather computer intensive as they work on a microscopic level, tracing the individual electrons.

One approach~\cite{But17} to this inverse problem is to use a CoREAS calculation as a template from which the emission amplitudes from the different shower-slices is stored. The emission from other showers is calculated from the template by simply adjusting the height-dependence of the weighting factor of the shower slices extracted from the template shower. Although very promising, several details still need further attention before this procedure can be applied.

As an alternative approach to solving the posed inverse problem we have developed a code that semi-analytically calculates the radio footprint of an extensive air shower for an arbitrary longitudinal structure of the electric current density in the shower front.
The analytic calculation uses a negligible amount of computer time, it is about 4 orders of magnitude faster~\cite{Sch16b} than a microscopic calculation (at E=$10^{16}$~eV), and, most importantly, does not suffer from random shower-to-shower fluctuations. Therefore it can be used in a chi-square minimization procedure to obtain the longitudinal structure that best reproduces the measured footprint.

For the present analytic code constructions similar to those that have been developed for the EVA-code~\cite{Wer12} are used to obtain the radiation fields from the Li\'{e}nard-Wiechert potentials.
Like EVA, the present code accounts for the proper retardation effects.
In EVA the parametrization of the plasma cloud is obtained from a Monte-Carlo simulation of a shower to be able to account for shower-to-shower fluctuations to the full extent. In the present code, however, we want to eliminate all shower-to-shower fluctuations and the shower evolution, including the structure of the plasma cloud, is parameterized. In this respect the model is similar to the MGMR model~\cite{Sch08} with the important difference that here the radial extent of the plasma cloud is taken into account. For this reason we named it \SName. Charge-excess radiation and a realistic index of refraction of air are also taken into account.

It is known from shower universality that the largest shower-to-shower fluctuations occur in the longitudinal shower evolution whereas the structure of the plasma cloud, like lateral extension and thickness, shows hardly any shower-to-shower fluctuations.
In order to keep the number of parameters manageable we have therefore adopted the approach where we use a generic parametrization of the structure of the plasma cloud where the dependence of the plasma-cloud integrated currents on the height in the atmosphere is left free since this is what we want to extract from the radio footprint.
The parametrization of the plasma cloud is inspired by the results of Monte-Carlo shower simulations and discussed in \secref{Modelling}. To validate the parameterizations we have verified in \secref{Compare} that the results of the \SName\ calculations for the radio footprint agree sufficiently well with those from microscopic CoREAS simulations.
Obtaining such a parametrization is an important part of the present work. As an interesting spin-off the code also allows to investigate which are the essential parameters of the plasma cloud for certain features of the radio pulse footprint. In this way -as an example- we noted that the temporal structure of the radio pulse, a strong peak followed by a very shallow undershoot, is strongly determined by the radial dependence of the pancake thickness, see \secref{Compare}.

Since the present code, \SName, is supposed to facilitate an iterative approach to reproduce a measured radio footprint much attention was given to its numerical stability and its calculational speed.
The code, \SName, is available from the authors upon request.

\section{Modeling Radio emission from EAS}\seclab{Modelling}

The currents and charges in the EAS are modeled as a plasma cloud with a parameterized density profile moving towards Earth at the speed of light, $c$. These currents are used to construct the retarded Li\'{e}nard-Wiechert potential from which the radiation fields are calculated. Here we closely follow the approach used in the EVA code.

To parameterize the charge cloud we introduce the shower-fixed coordinates $(t_s,x_s,y_s,h)$ where $t_s$ is the time when the shower front is at a distance $z_s=-t_s c$ from the ground (measured along the shower axis) and $(x_s,y_s)$ are transverse coordinates of a point in the shower plasma cloud at a distance $h$ behind the shower front.
The structure of the charge and current distributions are expressed through a four-current $j^\mu(t_s,x_s,y_s,h)$.
We use the notation where $\mu=0$ denotes the time (charge) components and $\mu=x,y,z$ the space (current) components of a four vector.  A particular point in the charge cloud is at a height of $\zeta=z_s+h$ as measured along the shower axis.

Following the usual notation where $t_r$ denotes retarded time the vector potential for an observer at a point $(t_o,x_o,y_o,z_o)$ in the shower plane, defined as the plane perpendicular to the shower axis going through the point of impact of the shower on the ground ($z_o=0$), is taken as
\beq
A^\mu(t_o,\vec{x_o})=\left. \int d^3 \vec{x}'\,\left|{dt_r\over dt_o}\right|\,{j^\mu(t_s,\vec{x}')\over L}\right|_{t_s=t_r} \;.
\eeq
Here $L$ is the optical path-length, $L=c(t_o-t_r)$, which for a homogeneous medium with a constant index of refraction $n$ is given by $L= nR=n|\vec{x_o}-\vec{x}'|$. Following the approach used in EVA~\cite{Wer12} we introduce ${\cal D}= L|dt_o/dt_r|$ as the retarded distance.
The vector potential can now be written more compactly as
\beq
A^\mu(t_o,\vec{x_o})=\int d^3 \vec{x}'\,{j^\mu(t_r,\vec{x}')\over {\cal D}} \;.
\eqlab{L-W}
\eeq
In the appendix, \secref{CE} and \secref{TC}, more details are given on the numerical calculation of the radiation fields from the vector potential \eqref{L-W}.

For simplicity the dependence of the index of refraction on the height in the atmosphere is taken as given by the Gladstone–-Dale~\cite{GD} relation,
\beq
n_{GD}=1+n_\rho \rho(z) \;,
\eeq
where $n_\rho$ is chosen such that the refractivity equals $dn=n-1=0.0003$ at ground level. This can be replaced by a more realistic dependence such as used in Ref.~\cite{Cor17} that takes the dependence of the refractivity on temperature and air humidity into account.

For the evaluation of the retarded distance the average value of the index of refraction will be used assuming the straight-line approximation for the photon path~\cite{Wer12},
\beq
n = (1/z)\int_0^z(1+n_\rho \rho(\zeta))\,d\zeta \;,
\eeq
which does not depend on the distance of the observer to the shower axis.
Thus to a good approximation (1:straight line trajectory; 2:index of refraction does not change with observer position) the retarded distance can be expressed as
\bea
{\cal D}&\stackrel{1}{=}& nR \left|{dt_o \over dt_r}\right| \stackrel{2}{=} nR\left.(1-n \vec{\beta}\cdot \hat{n})\right|_{\mbox{ret}}
 \nonumber \\
 &=& (t_o-t_r) - n^2\beta(h-\beta t_r)
 \nonumber \\
 &=& n\sqrt{(-\beta t_o +h)^2 + (1-\beta^2 n^2)d^2}
 \eqlab{denom} \;,
\eea
for a moving point charge with velocity $\beta c$ and an observer at time $t_o$ at a distance $d=\sqrt{x_o^2+y_o^2}$ from the shower axis at $z_o=0$.

The four current is parameterized as
\beq
j^\mu(t_s,x_s,y_s,h)={w(r_s) \over r_s} \, f(h,r_s) \, J^\mu(t_s) \;,
\eqlab{DefCloud}
\eeq
where $r_s=\sqrt{x_s^2+y_s^2}$, and the functions $w$ and $f$ are normalized according to $\int w(r)\, dr=1$ and $\int_0^\infty f(h,r)\, dh=1\; \forall\, r$. In this way $J^\mu(t_s)$ is the charge and current at time $t_s$ integrated over the complete plasma cloud of the EAS at a fixed time.

\subsection{Parameterizations}

In the parameterizations of the charge cloud we first concentrate on the structure of the charge/current cloud under fair weather circumstances.  The effect of atmospheric electric fields is considered in a following step. It should be noted that these fields will primarily change the magnitude and orientation of the transverse electric currents, $J^\mu(t_r)$, however, as shown in Ref.~\cite{Tri15}, these fields will also affect the structure of the current cloud, in particular $f(h)$, see \eqref{DefCloud}.

Most of the plasma-cloud parameters are obtained through a comparison with the results of a Monte-Carlo simulation of of an extensive air shower. Based on shower universality we use a single shower in order not to have complications of averaging the results of two showers with different values for $\Xmax$. As noted in the following discussions, for some case we have preferred a CONEX-MC simulation, the same that lies at the basis of the EVA calculations~\cite{Wer12}, for ease of extracting more detailed information of the shower structure while for others, where the atmospheric electric field is important, we used CORSIKA.

\subsubsection{Fair-weather conditions}

The spatial extent of the charge cloud is modeled as given in \eqref{DefCloud} where the functions $w$ and $f$ are normalized to unity.
%
Under fair-weather conditions the radial dependence of the transverse current is parameterized as
\beq
w(r_s)=N_w\,\xi (\xi+1)^{-2.5} \;,  
\eqlab{Def-w}
\eeq
with $\xi=r_s/M_0$ introducing the Moliere radius $M_0$ as a scaling factor and where $N_w$ is chosen such that $\int w(r)\, dr=1$. Note that $w(r_s)$ corresponds to $r_s$ times the  Nishimura-Kamata-Greisen   function for $s=2$~\cite{NKG}. In \figref{w} we compare the results of CONEX-MC~\cite{CONEX} (the Monte-Carlo version of CONEX) with the parametrization \eqref{Def-w} using the parameters given in \appref{Param}. 

\begin{figure}[h]
 \includegraphics[width=0.40\textwidth]{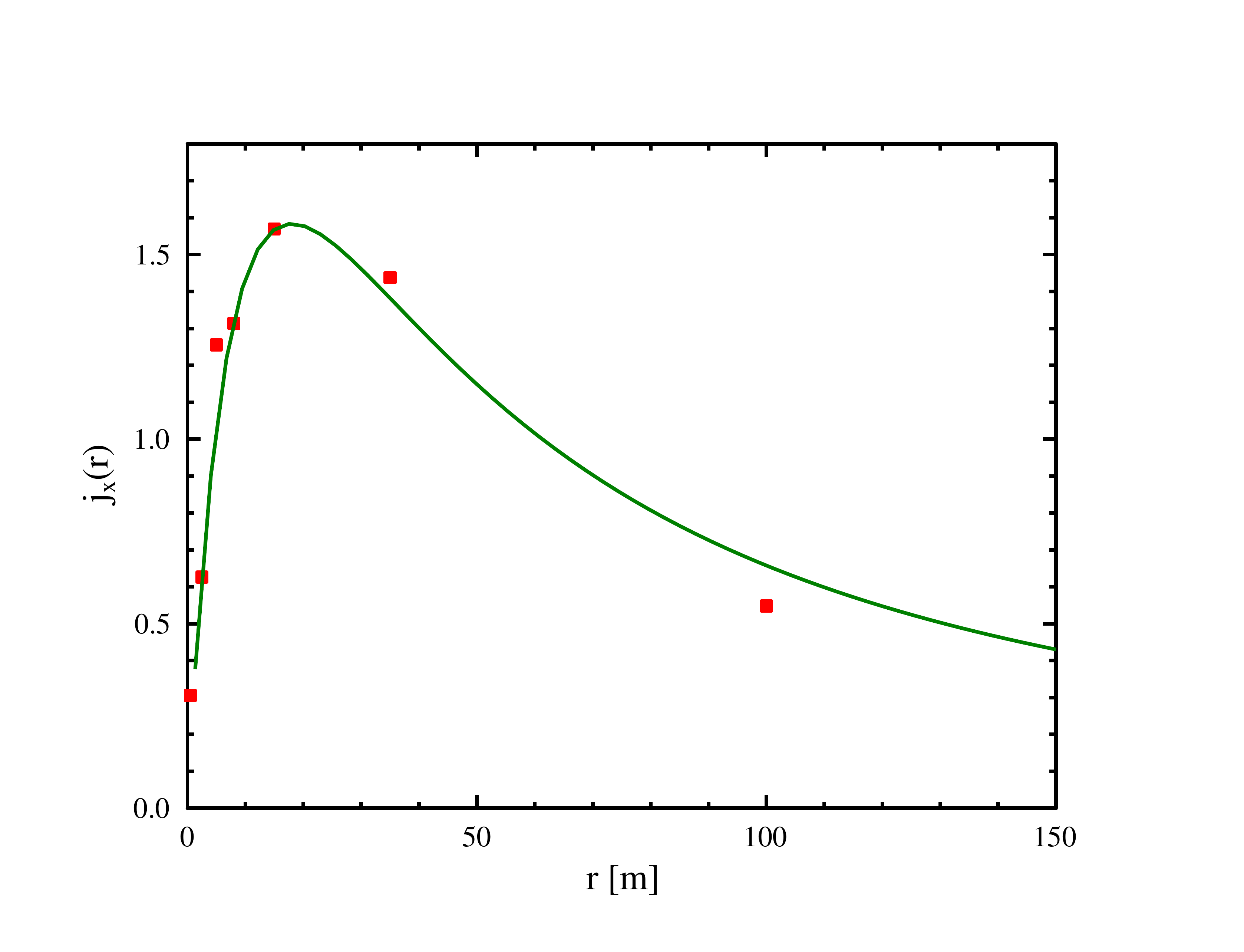}
 \caption{Radial dependence of the shower transverse current, \eqref{Def-w} (line) is compared with CONEX-MC simulations (dots).\figlab{w}}
\end{figure}

The current density at a distance $h$ behind the shower front is parameterized as
\beq
f(h,r_s)=N_f\, {\eta \over e^{\sqrt{\eta}} + 1}\;,  
\eqlab{Def-f}
\eeq
where $\eta={h/ \lambda}$.  The norm, $N_f$ is chosen such that $\int_0^\infty f(h,r_s) dh=1$ for all values of $r_s$. The pancake-thickness scaling parameter
\beq
\lambda(r_s,|F|)= \Lambda(r_s) \, \alpha(|F|) \;, 
\eqlab{Def-lam}
\eeq
is factorized in a dependence on distance to the shower axis, $r_s$,
\beq
\Lambda(r_s)= \max[\Lambda_0, \Lambda_1 {r_s\over r_1} ] \quad \;,  
\eqlab{Def-lamr}
\eeq
and a scale parameter $\alpha(|F|)$ that depends on the net force acting on the electrons, $|F|$, and is specified in more detail in \eqref{Def-alphaE}.
The radial dependence of the pancake thickness is such that near the shower core we have $\Lambda(0)=\Lambda_0$ while increasing linearly at larger distance where at a distance of $r_s=r_1=$100~m from the core we have $\Lambda(r_1)=\Lambda_1$.

In principle the pancake-thickness scale parameter $\alpha$ may also depend on penetration depth. We observe however that for fair-weather showers the radio footprint as well as pulse shape (discussed in the following chapter) show very little sensitivity to such a height dependence and are determined almost solely by the pancake thickness at the shower maximum. We thus ignore this possible dependence.

In \figref{jrh} the parametrization of the pancake structure, using parameter values from \appref{Param}, is compared to the results of a CONEX-MC~\cite{CONEX} simulation for a fair-weather shower at the shower maximum.

\begin{figure}[h]
 \includegraphics[width=0.40\textwidth]{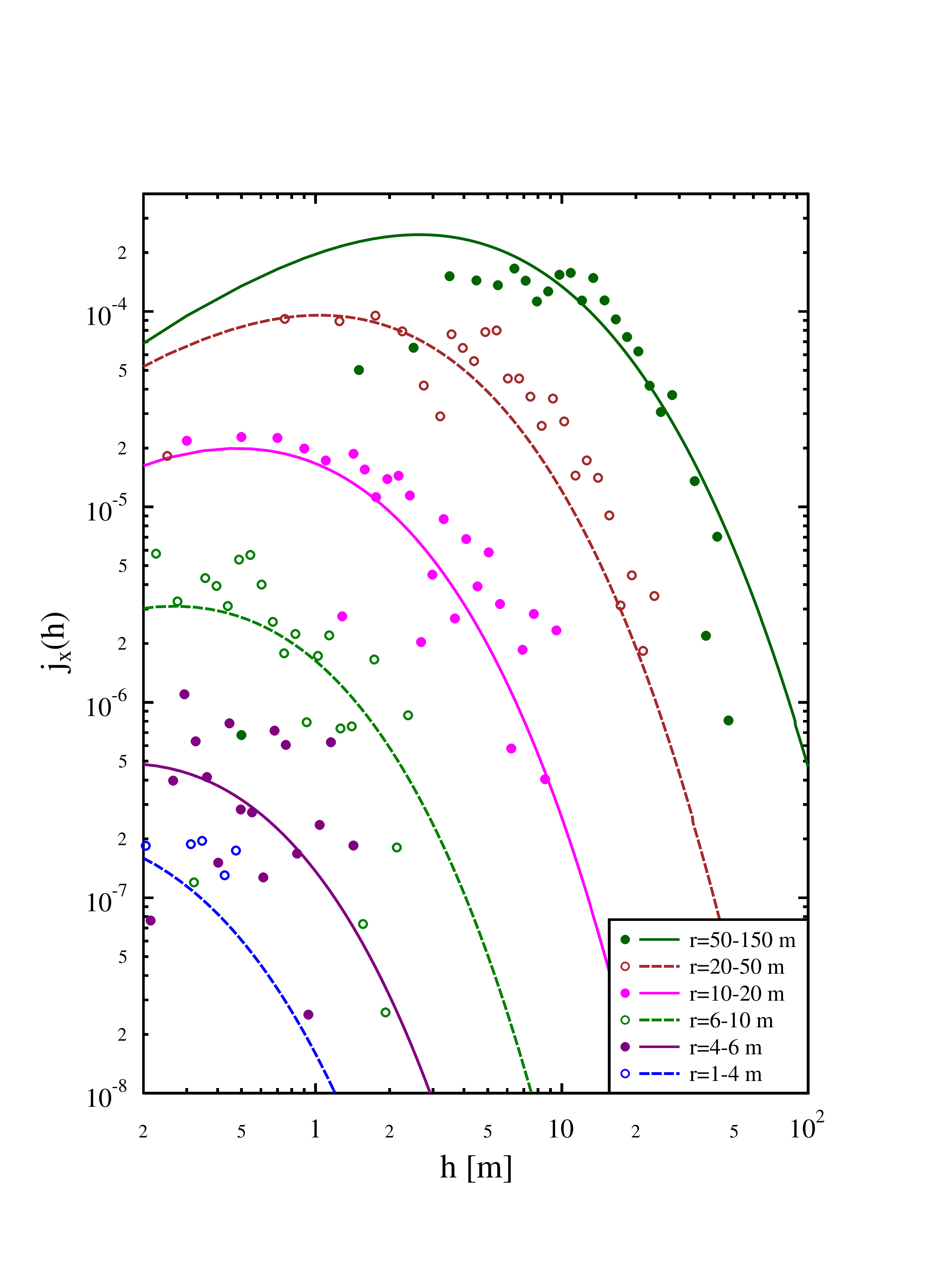} 
 \caption{The parameterized dependence of the current density in the plasma (curves), \eqref{Def-f}, as a function of distance to the shower front at several distances from the shower axis (at the height of the shower maximum) is compared to the results of CONEX-MC (dots). The data at the different distances from the core are off-set by factors of powers of 4. 
 \figlab{jrh}}
\end{figure}

The parametrization  of the longitudinal shower profile for the charge excess and the transverse current in the plasma cloud for a shower under fair weather conditions is based on the  Gaisser-Hillas formula~\cite{Gai77} for the dependence of the number of charged particles on $\Xrh$, the penetration depth,
%
\beq
N_{c}(\Xrh)= \left({ \Xrh-X_0 \over \Xmax -X_0}\right)^{(\Xmax -X_0)/ \gamma} e^{(\Xmax - \Xrh)/\gamma}
\eqlab{Def-Nc}
\eeq
where $\gamma$ is a parameter controlling the width of the distribution and $X_0$ the reference point. The transverse current, see \eqref{DefCloud}, is obtained by multiplying the number of charged particles with the drift velocity,
\beq
\vec{J}_\perp(t_s)=N_{c}(\Xrh) \,\vec{u}_\perp(\Xrh)
\eqlab{Def-It}
\eeq
where the induced transverse drift velocity is denoted as $\vec{u}_\perp$.

The drift velocity will increase with increasing forces acting on the charges, however, for large forces, as we will encounter during thunderstorm conditions, one should be careful to take into account that the velocity of the particles does not exceed the speed of light. Following the arguments given in Ref~\cite{Tri15} we take,
\beq
\vec{u}_\perp(\Xrh)=c \vec{\upsilon} /\sqrt{1 +\upsilon^2/\upsilon_0^2} \;,
\eqlab{Def-u}
\eeq
where the parameter $\upsilon_0$, as discussed in a later section, is taken such that for fair weather conditions one is still in the regime where the drift velocity scales linearly with the Lorentz force. When not too large, the drift velocity is proportional to the force acting on the plasma charges,
\beq
\vec{\upsilon}(\Xrh)= {\vec{F}_\perp \over F_\beta}\, {(1+a_t)^2\Xrh\, \sqrt{\Xmax\,X_v}\over (\Xmax + a_t\,\Xrh)^2 } \;,
\eqlab{Def-sigma}
\eeq
where $F_\beta$ represents a friction constant, $a_t=2$, and a normalization constant of $X_v=500$~g/cm$^2$ is used.
The last factor in \eqref{Def-sigma} takes into account the fact that the drift velocity depends on the penetration depth in the atmosphere. At low altitudes the drift velocity decreases due to increased density and at high altitudes, early in the shower development, the energy of the particles in the plasma is enormous, and for this reason their sidewards drift is small.
For fair weather conditions only the Lorentz force is acting, $\vec{F}_\perp=e \vec{v}_s \times \vec{B}$, and is constant along the shower. Here $\vec{v}_s$ is along the shower with magnitude $c$ and $\vec{B}$ is the magnetic field.
In  \figref{Jxz} the parametrization for the drift velocity is compared with the results of a CONEX-MC calculation for a vertical shower where a geomagnetic field of 56~$\mu$T is used at an angle of $63^\circ$ with the horizontal. Using this figure the values for $F_\beta$ and $a_t$  in \eqref{Def-sigma} are deduced where their values are given in \appref{Param}.
The arguments for introducing $\upsilon_0$ are discussed in a following section.

The longitudinal current due to charge excess is defined as,
\beq
J_Q(z)=N_{c}(\Xrh)  \, \rho_c(\Xrh) \;,  
\eqlab{Def-IQ}
\eeq
where,
\beq
\rho_c(\Xrh)=J^0_Q\,{1+a_c \over a_c+ \Xmax/\Xrh} \;,   
\eqlab{Def-rhc}
\eeq
models the dependence of the charge excess fraction on penetration depth with $a_c=0.5$. This parametrization is compared with the results of a simulation in \figref{Jxz} while $J^0_Q$ is a normalization constant.

\begin{figure}[h]
 \includegraphics[width=0.40\textwidth]{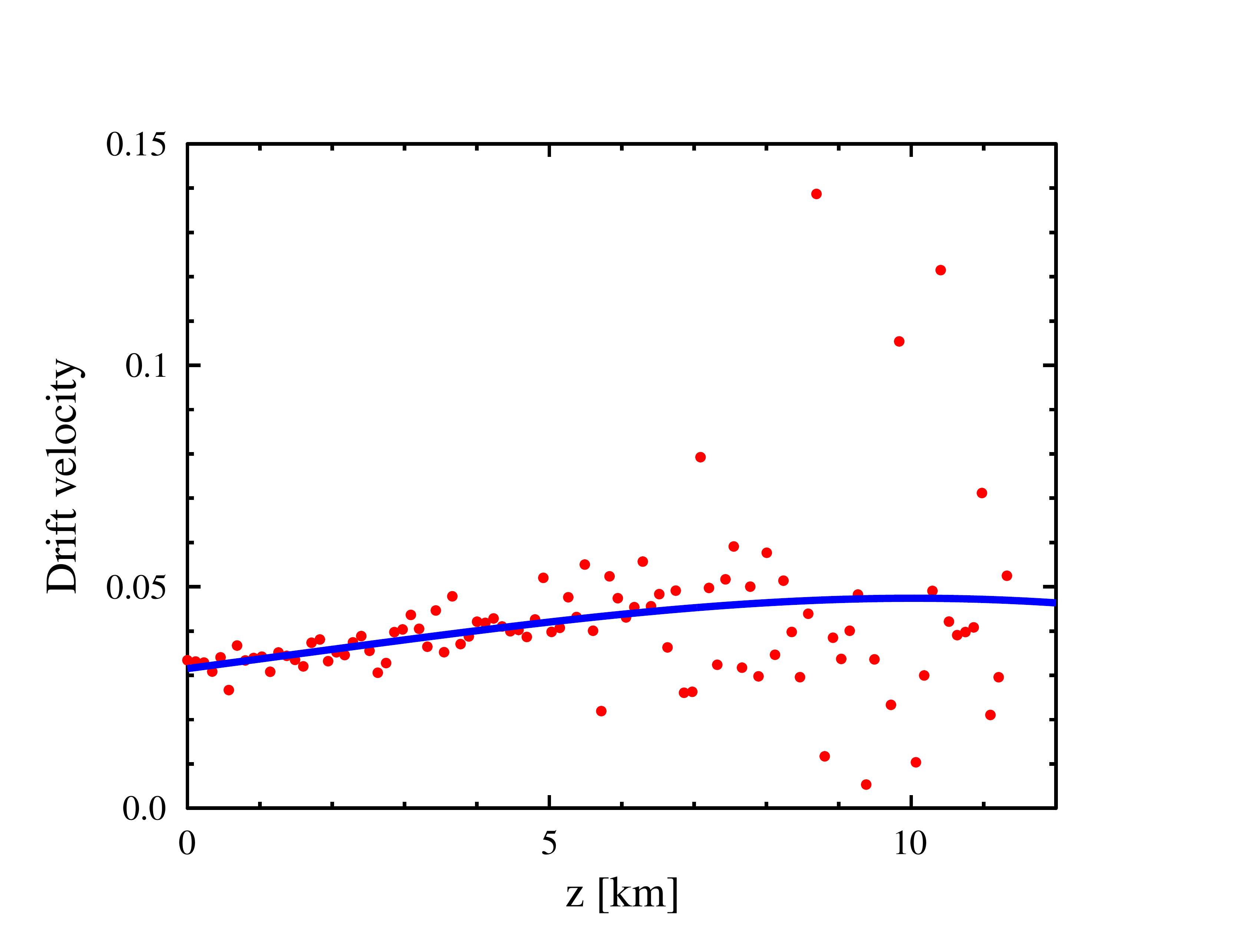} \includegraphics[width=0.40\textwidth]{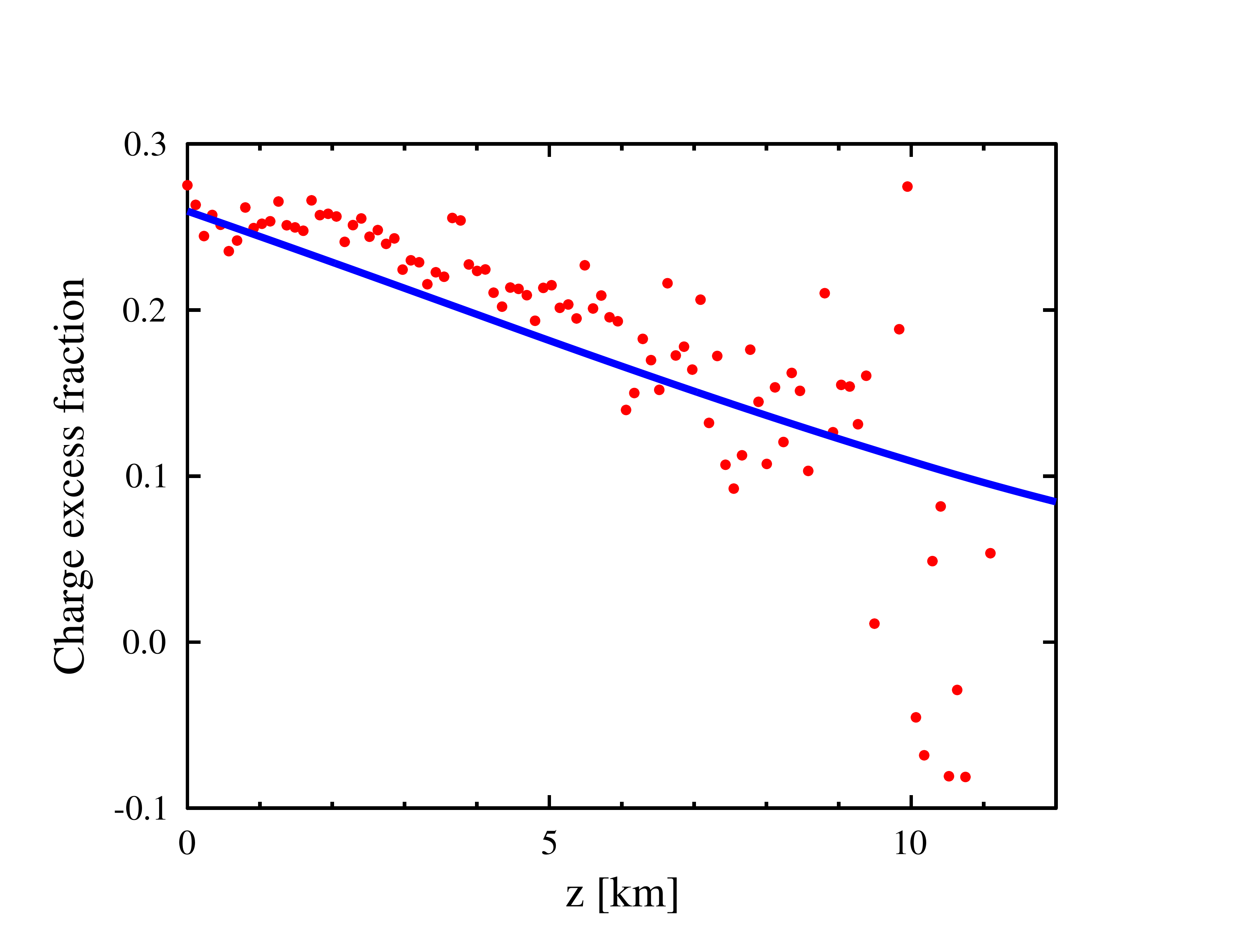}
 \caption{Comparing the height dependencies, $z_s$, of the transverse drift velocity and the charge excess  for the parameterized shower and the one simulated in CONEX-MC. \figlab{Jxz}}
\end{figure}


For an inclined shower at an angle $\theta_s$, ignoring the curvature of Earth, we have $H=z\cos{\theta_s}$ and the height dependence of the atmospheric penetration depth is taken as
\beq
\Xrh(z)= \left( a+b\, e^{-H/c} \right) /\cos{\theta_s} \;,   
\eqlab{Def-Xrh}
\eeq
where $H$ denotes the height above ground and where the constants $a,b,c$ depend on height as given in \tabref{Atmosphere} for the U.S. standard atmosphere~\cite{USSA} which are the same as used in CORSIKA~\cite{CORSIKA}.

\begin{table}[!ht]
\caption{Parameters used for the air-density profile\tablab{Atmosphere}}
\begin{tabular}{h{3}||d{7}|d{5}|d{5}|}
\hline
\multicolumn{1}{c||}{$H$ [km]} & \multicolumn{1}{c|}{a [g/cm$^2$]} & \multicolumn{1}{c|}{b [g/cm$^2$]} & \multicolumn{1}{c|}{c [m]} \\
\hline
10-20 & 0.61289 & 1305.5948 & 6361.4304\\
4-10  & -94.919 &  1144.9069 &  8781.5355 \\
0-4  & -186.555305 &  1222.6562 &  9941.8638\\
\hline
\end{tabular}
\end{table}


\subsubsection{Thunderstorm conditions}

In the presence of thunderclouds the air shower will generally evolve through areas in the atmosphere where there are large electric fields. These will significantly alter the currents in the shower front~\cite{Sche15,Tri15}. It is precisely these current we want to determine from the radio footprint. In leading order the electric fields will change the magnitude and direction of the induced drift velocities,
\beq
\vec{F}_\perp=e \left[ \vec{v} \times \vec{B} + \vec{E}_\perp \right] \;, \eqlab{FE}
\eeq
in \eqref{Def-sigma}.
We will assume that the strength of the component of the field parallel to the shower is below the limit where secondary electron avalanches are formed. For these strengths the number of particles in the shower is not significantly affected~\cite{Tri15} and we thus can ignore it. The component of the field perpendicular to the shower directly influences the current.

An important secondary effect of the electric field is to increase the pancake thickness~\cite{Tri15}. Since the particles in the shower front are constantly being regenerated from the more energetic ones that drive the shower, the pancake structure is affected only at those heights where the field acts. The structure of the plasma cloud thus is hardly determined by its structure at larger heights and it thus can be said that there are no memory effects. This has been checked with Monte-Carlo simulations.

\begin{figure}[h]
 \includegraphics[width=0.40\textwidth]{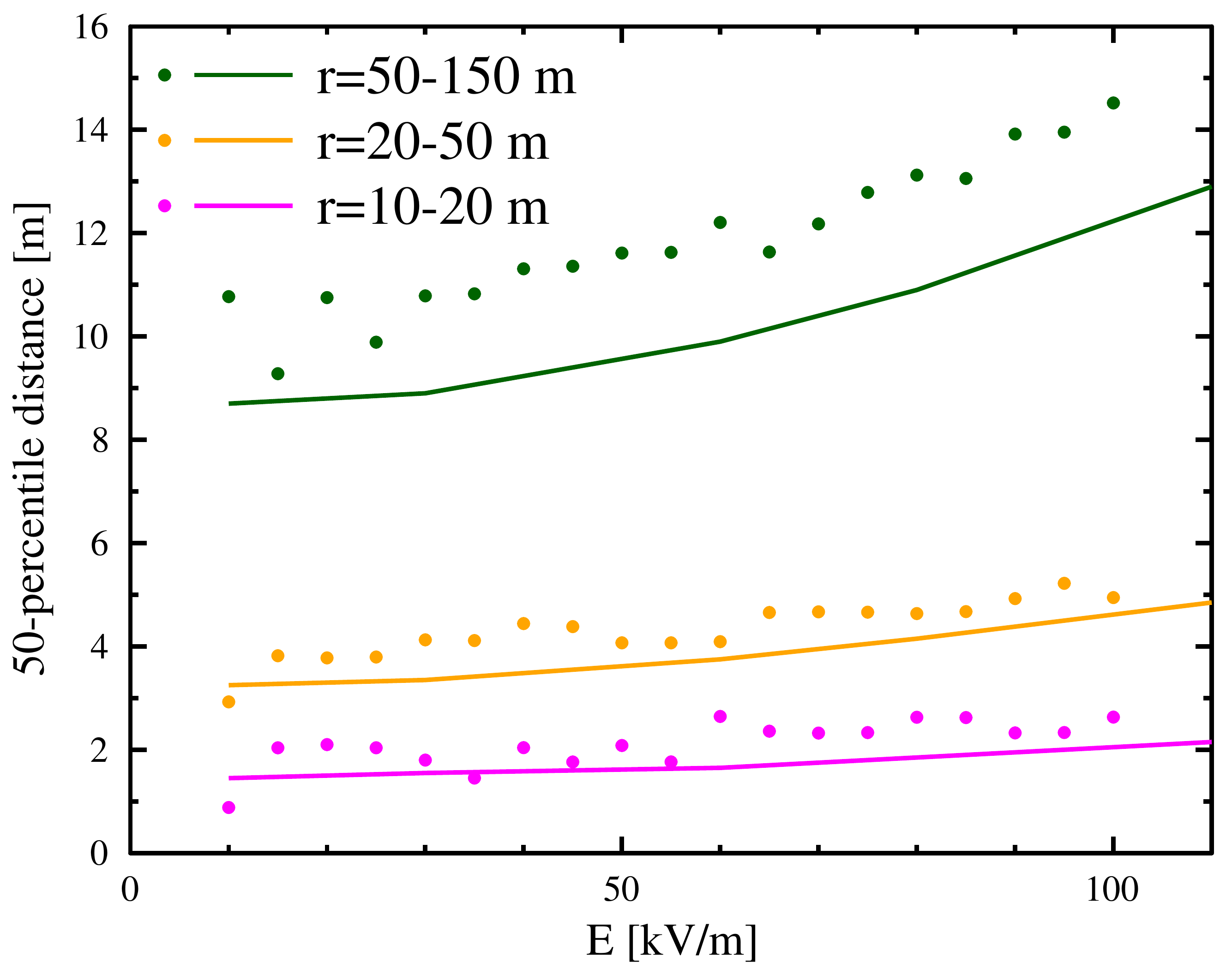}
 \caption{The 50 percentile distance of particles behind shower front for three intervals in radial distances from axis as calculated in CORSIKA (dots) and with the present parametrization (curves). \figlab{E-pancake}}
\end{figure}

Monte-Carlo simulations show that the distance from the shower front that contains 50\% of the number of charged particles near the shower maximum depends  quadratically on the transverse force, see \figref{E-pancake}. This effect we have parameterized through a dependence of the pancake-thickness scaling factor $\alpha$, see \eqref{Def-lam}, on the strength of the transverse force, $|F|$, acting on the particles in the plasma cloud,
\beq
\alpha(|F|)= 1 
+ a_E \, \left| {\vec{F}_\perp \over 100\, {\rm [keV/m]}} \right|^2 \;.  
\eqlab{Def-alphaE}
\eeq
The parameter $a_E$ is adjusted to reproduce the median trailing distance behind the shower front as obtained from CORSIKA simulations, see \figref{E-pancake}. Using the parameter values given in \appref{Param} a good agreement is obtained.

\section{Comparison with CoREAS simulations}\seclab{Compare}

\begin{figure}[ht]
    \centering
    \begin{subfigure}[b]{0.35\textwidth}
        \includegraphics[width=\textwidth]{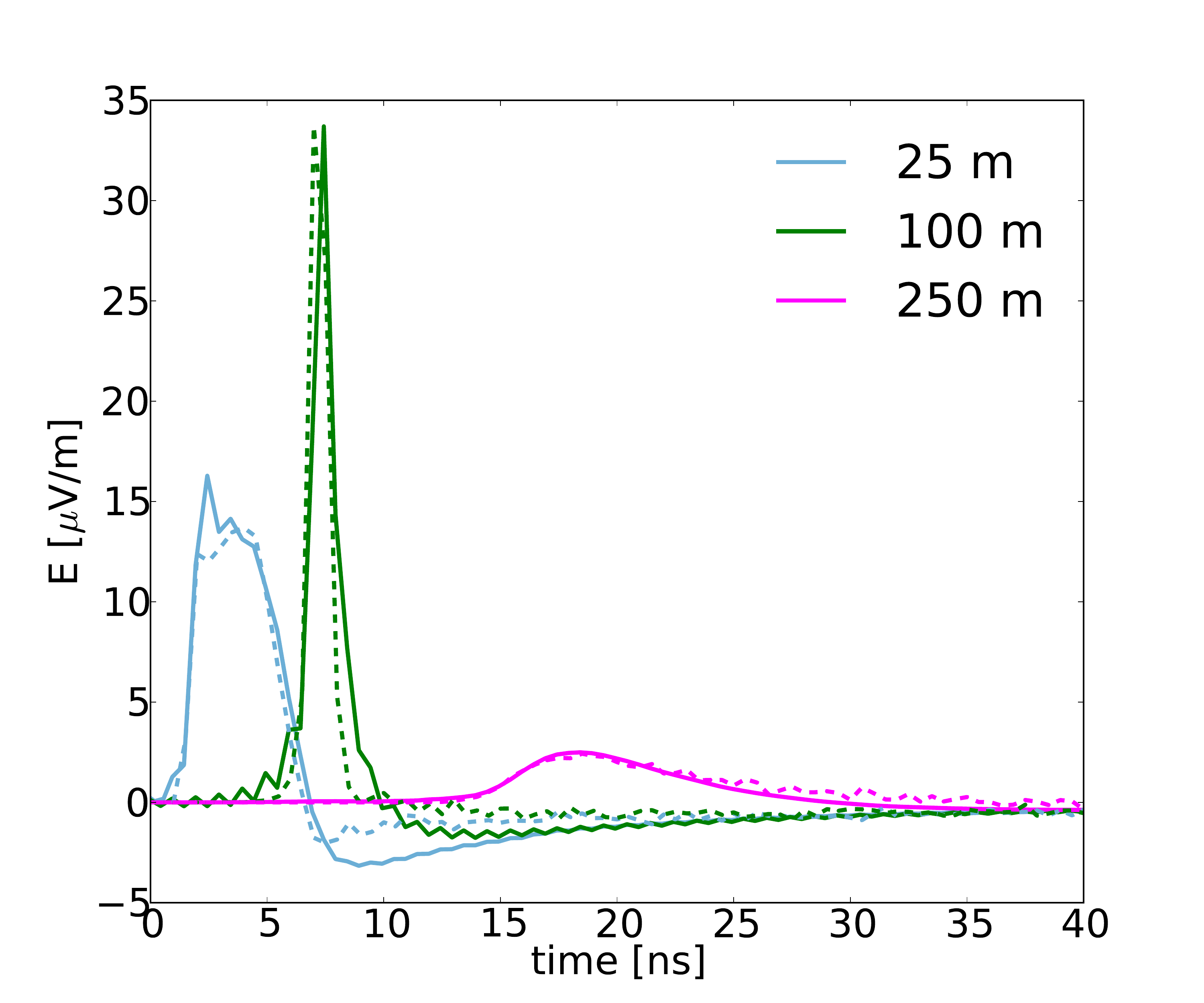}
        \caption{} \figlab{Pulse35}
    \end{subfigure} \\
    ~ 
    \begin{subfigure}[b]{0.35\textwidth}
        \includegraphics[width=\textwidth]{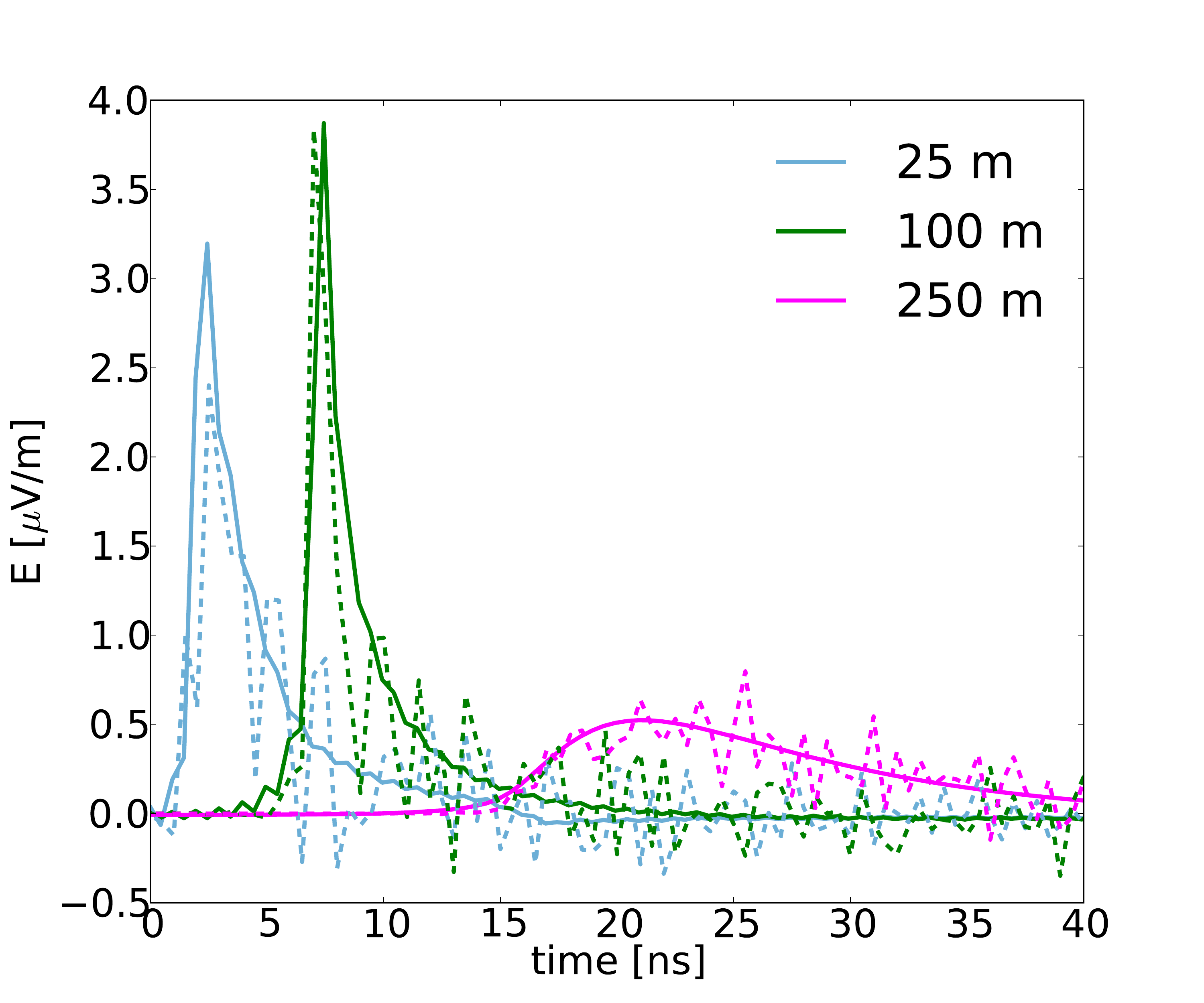}
        \caption{} \figlab{Pulse36}
    \end{subfigure}
\caption{\small  Pulse shapes as calculated with \SName\ (full lines) and compared to CoREAS (dashed lines) for a vertical shower with $\Xmax=540$ g/cm$^2$.
Fig.~\subref{fig:Pulse35}) transverse current, Fig.~\subref{fig:Pulse36}) charge excess.  }\figlab{Pulses}
\end{figure}

Having fixed the the parameters of the plasma cloud as given in \appref{Param} we need to verify that the used parametrization is sufficiently detailed such that the radio footprint from \SName\ agrees with the CoREAS results to a reasonable accuracy.

As a first step we compare in \figref{Pulses} the unfiltered pulses separately for the transverse current and the charge excess contributions at various distances to the core with the results of a CoREAS simulation. The calculations are performed assuming fair-weather circumstances for a vertical shower with $\Xmax=540$~g/cm$^2$.
At this point it is worthwhile to note that -in principle- there can be an additional contribution to the emitted radiation due to an induced dipole distribution that is co-moving with the shower front~\cite{Sch08}. The geometry of its radiation pattern is very similar to that of geomagnetic radiation, however the pulse is very elongated in time. We have calculated such a contribution and seen that best agreement with the microscopic calculation is obtained by setting this contribution to zero. This implies that the net displacement of electrons and positrons in the shower front is vanishingly small. This is in line with the conclusions reached in Ref.~\cite{Alv14}. Another interesting point is that the typical structure of the pulse with a large positive peak followed by a long negative tail is intimately linked to the radial dependence of the pancake thickness. When taking a less pronounced increase with distance (i.e. decreasing the value of $\Lambda_1$ in \eqref{Def-lamr}) the negative tail gets shorter and more pronounced. For the case in which $\lambda$ is independent of distance to the core a clear bi-polar pulse is obtained.

\begin{figure*}[t]
 \includegraphics[width=0.94\textwidth]{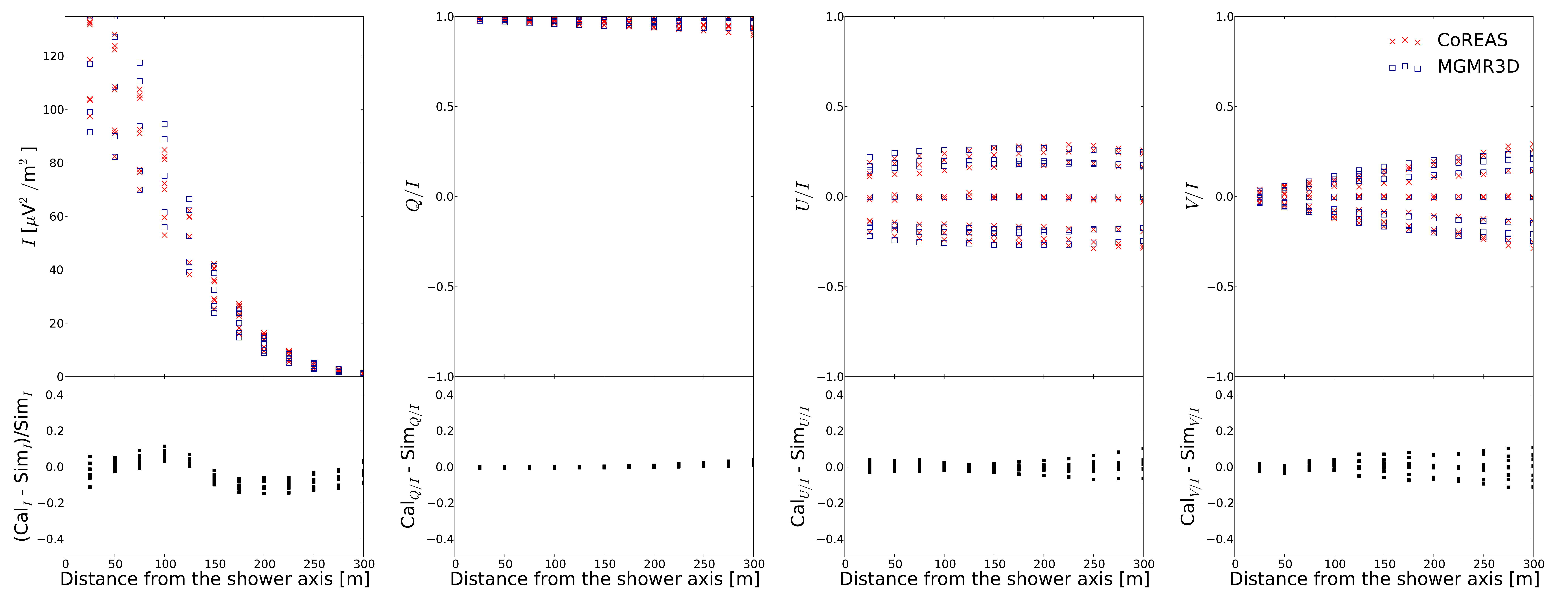}
 \caption{The Stokes parameters calculated with \SName\ (blue squares) are compared with those from a microscopic CoREAS calculation (red crosses) for a shower with $\Xmax=540$~g/cm$^2$. The pulses are filtered between 30 and 80~MHz as is realistic for LOFAR.  \figlab{FW-stokes-H}}
\end{figure*}

\begin{figure*}[t]
 \includegraphics[width=0.94\textwidth]{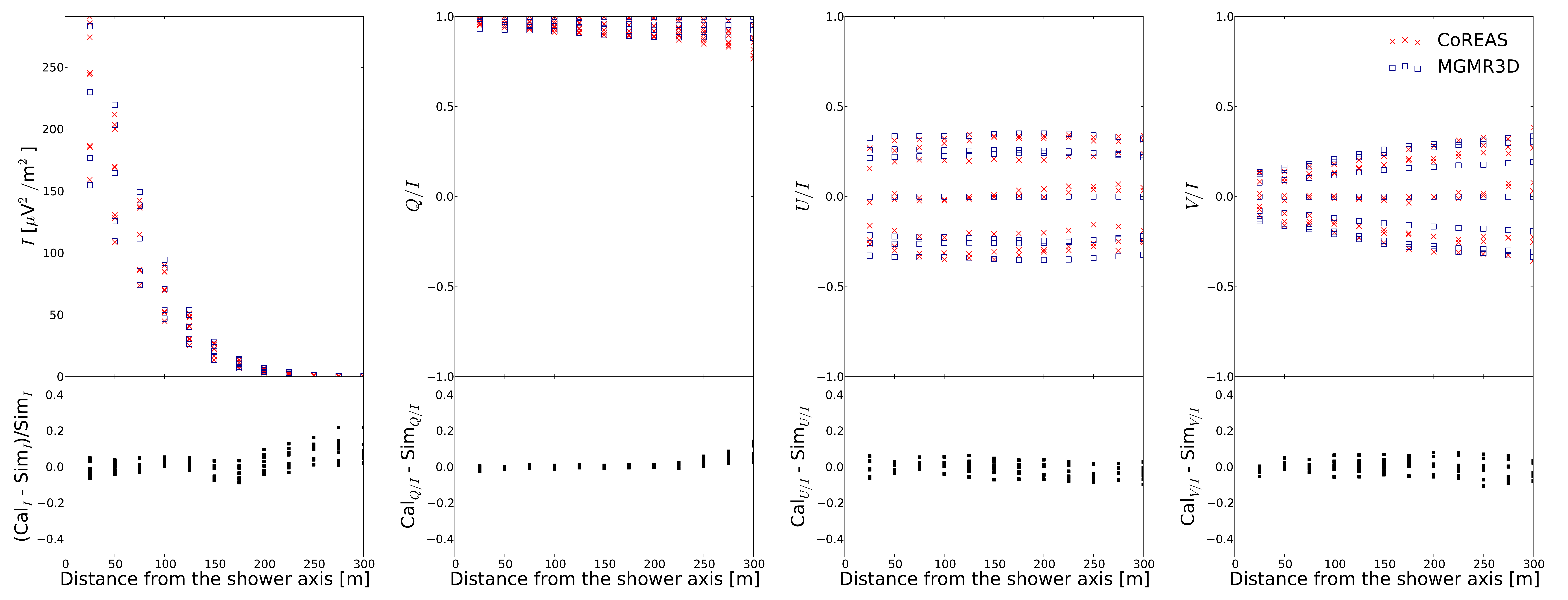}
 \caption{Same as \figref{FW-stokes-H} for a shower with $\Xmax=690$~g/cm$^2$. \figlab{FW-stokes-L}}
\end{figure*}

\subsection{Fair-weather footprint}

We will investigate the radio footprint of an air shower using Stokes parameters since these capture the complete polarization structure of the radio pulse. Because the objective of the present work is to develop a scheme that can be used to ease the interpretation of data, we construct the Stokes parameters with the LOFAR~\cite{Haa13} cosmic-ray experiment~\cite{Sche14} in mind. This implies filtering the signal to the 30 -- 80~MHz band. In terms of the sampled pulse in the polarization direction $p$, where the complex voltage of the $i^{th}$ sample are denoted as ${\cal E}_{i,p}=E_{i,p} + i\hat{E}_{i,p}$, the Stokes parameters can be expressed as
\begin{eqnarray}
I&=&{1\over n} \sum_0^{n-1} \left( |{\cal E}|^2_{i,\vB} + |{\cal E}|^2_{i,\vvB} \right) \nonumber\\
Q&=&{1\over n} \sum_0^{n-1} \left( |{\cal E}|^2_{i,\vB} - |{\cal E}|^2_{i,\vvB} \right) \nonumber\\
U +iV&=&{2\over n} \sum_0^{n-1} \left( {\cal E}_{i,\vB} \;  {\cal E}_{i,\vvB}^* \right) \;.
\eqlab{Stokes}
\end{eqnarray}
$\hat{E}_{i,p}$ is the Hilbert transform~\cite{Sche14} of the real measured voltage $E_{i,p}$. The polarization directions $p$ are taken along  $\hat{e}_\vB$ and $\hat{e}_\vvB$ which are by construction perpendicular to the propagation direction of the photon (in very good approximation).
We sum over the whole trace.
The linear-polarization angle with the $\vB$-axis, $\psi$, can be calculated directly from the Stokes parameters as $\psi={1\over 2} \tan^{-1} (U/Q)$. The relative amount of circular polarization is given by $V/I$.

A comparison of the Stokes parameters~\cite{Sche14,Tri16} with the results of a microscopic CoREAS simulation is presented in \figref{FW-stokes-H} for the case of a relatively small value, $\Xmax=540$~g/cm$^2$ and in \figref{FW-stokes-L} for a larger value, $\Xmax=690$~g/cm$^2$. A simple geometry is used with a vertical shower and a horizontal magnetic field with $B=40~\mu$T inducing a transverse force of $F_\perp=12$~keV/m. The Stokes parameters are calculated for fair weather circumstances for a star-shaped layout of antennas with the center on the shower axis and arms at 0$^\circ$, 45$^\circ$, and 90$^\circ$.  The lower panels in the figures show the difference between the microscopic and macroscopic calculation.
The intensities of the radio emission as calculated from \SName, filtered between 30 -- 80~MHz, agree very well with  those of the CoREAS simulations.
For almost pure polarization in the $\vB$ direction one obtains $Q/I \approx 1$. Due to the effects of the small contribution of charge excess radiation the polarization angle deviates slightly from the $\vB$ direction which gives rise to $U/I \neq 0$ where the magnitude and the strength depends on the azimuthal orientation of the antenna with respect to the shower core. Likewise the circular polarization, expressed by $V/I$, is small and azimuth angle dependent. There is a good agreement in the polarization directions between \SName\ and CoREAS.

\begin{figure*}[t]
 \includegraphics[width=0.94\textwidth]{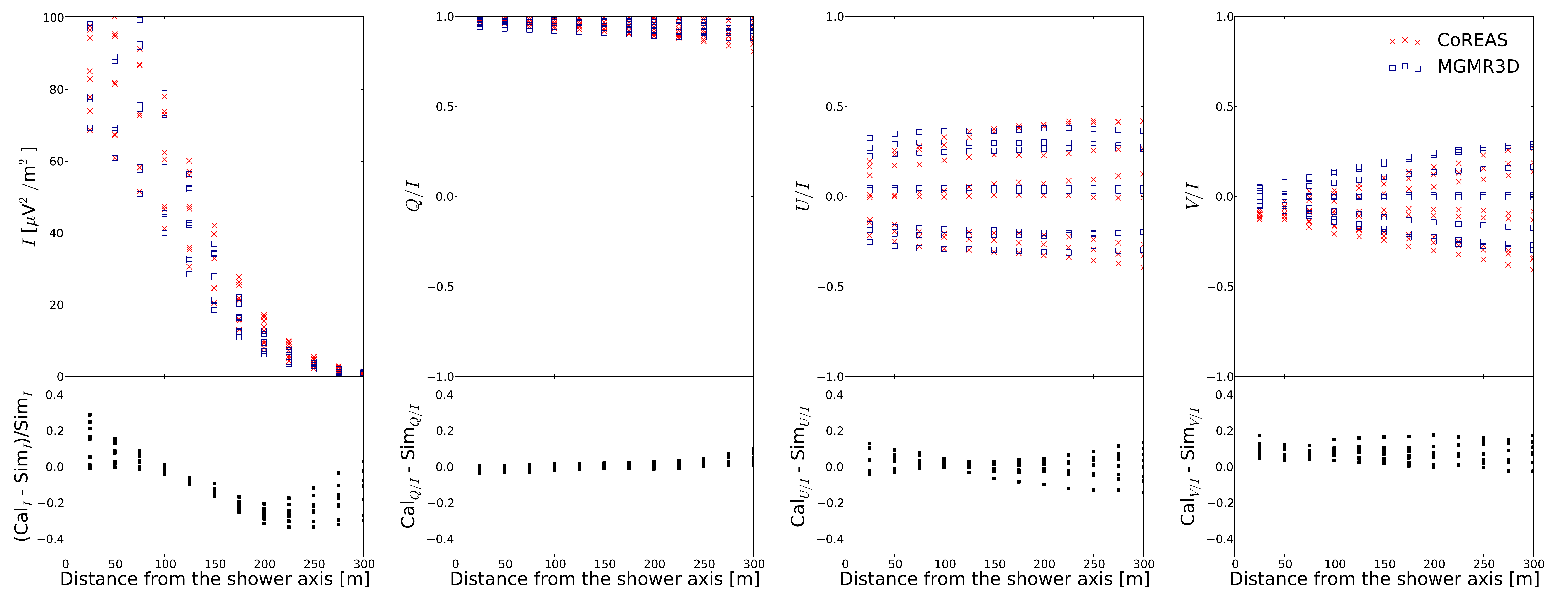}
 \caption{Fair-weather, $\Xmax=693$g/cm$^2$, inclined at 30$^\circ$. \figlab{FW-stokes-Incl}}
\end{figure*}

In \figref{FW-stokes-Incl} we present the results for a shower with a zenith angle of 30$^\circ$ and $\Xmax=693$~g/cm$^2$. The magnetic field is at an angle of $\alpha_{vB}=60$ degrees with the shower axis. In such a configuration care should be taken with the subtle effect that there is an additional component in the Lorentz force due to the drift velocity of the particles. This is taken into account by adding to the currents an additional component proportional to the component of the magnetic field parallel to the shower axis, $\vec{B}_\parallel$,
\beq
\vec{J}'_\perp=\vec{J}_\perp \times \vec{B}_\parallel / F_\beta \;.
\eqlab{BxvxB}
\eeq
As can be seen from \figref{FW-stokes-Incl} the intensity footprint calculated with \SName\ agrees rather well with the one calculated with CoREAS although there are some systematic differences. The reasons for these differences is not understood. Another surprising, and not understood (small) discrepancy between CoREAS and \SName\ is seen for $V/I$. Even with the correction from \eqref{BxvxB} the circular polarization near the core is vanishingly small in \SName\ while CoREAS shows a clear off-set.

\subsection{Footprint at higher frequencies}

\begin{figure*}[ht]
 \includegraphics[width=0.94\textwidth]{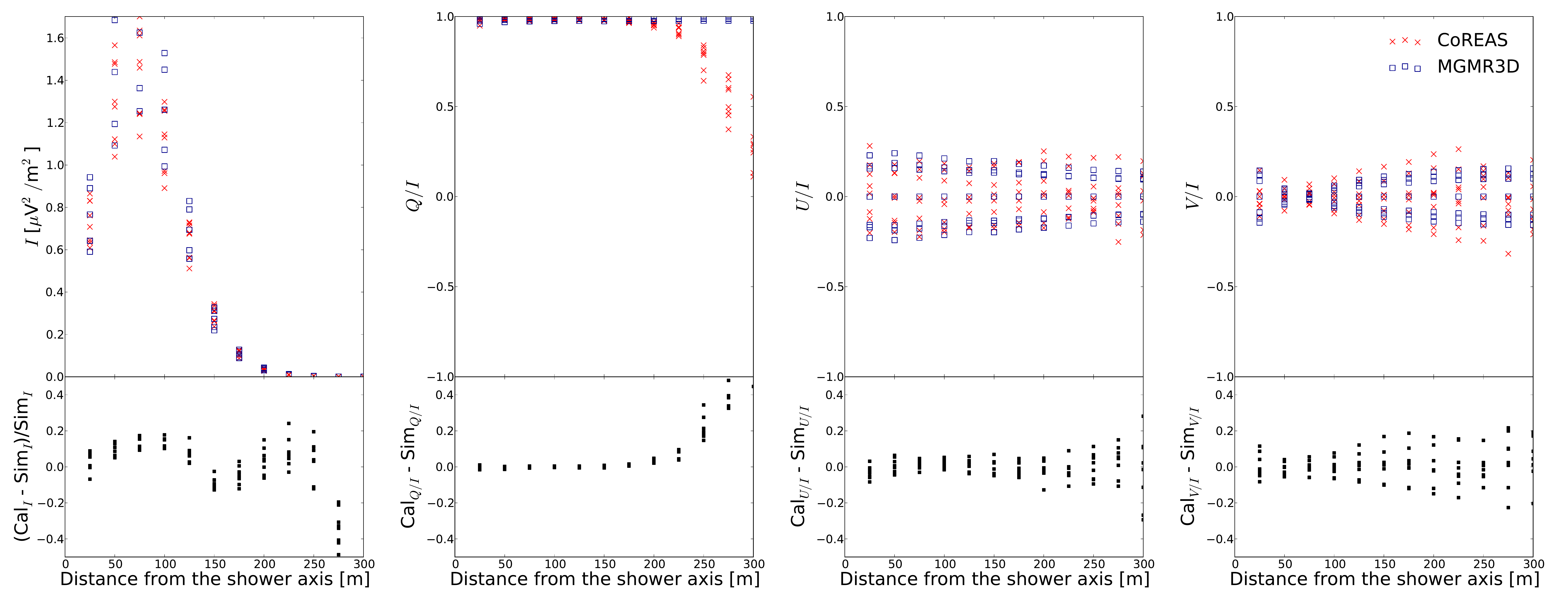}
 \caption{Same as \figref{FW-stokes-H} for the bandwidth of 100 -- 200~MHz. \figlab{FW-stokes-100}}
\end{figure*}

\begin{figure*}[ht]
 \includegraphics[width=0.94\textwidth]{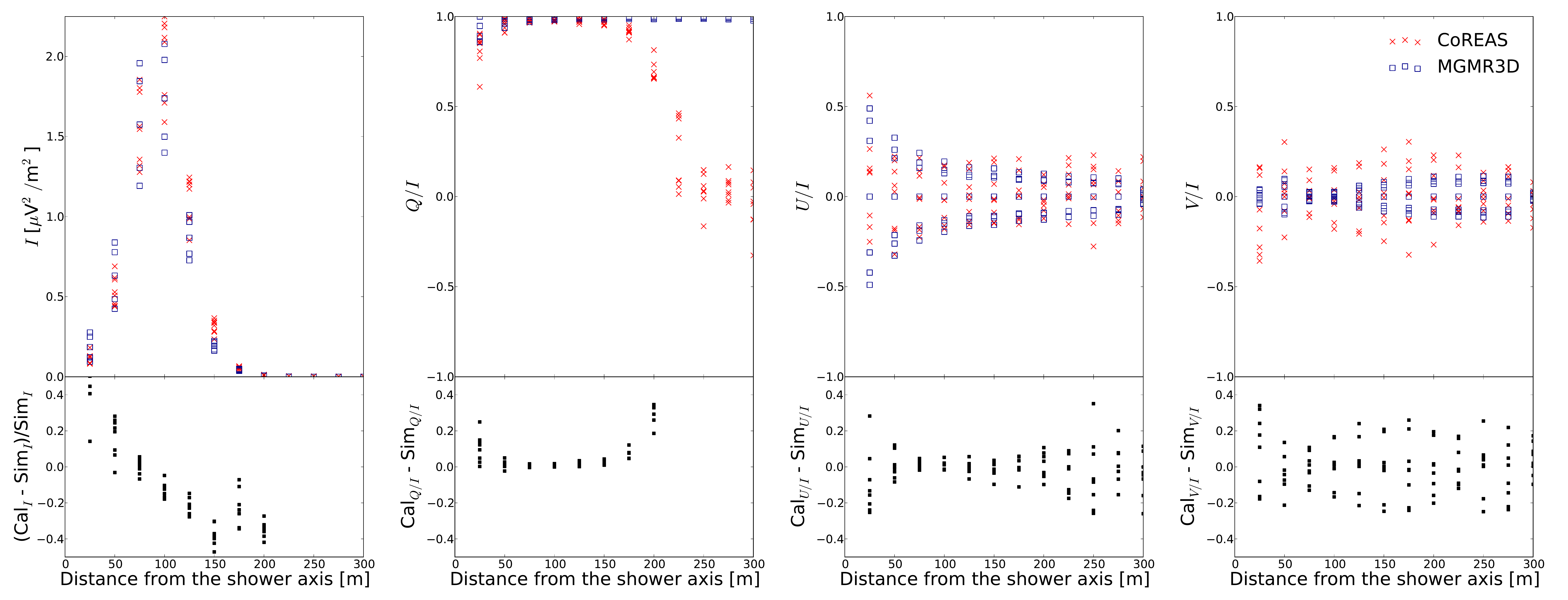}
 \caption{Same as \figref{FW-stokes-H} for the bandwidth of 200 -- 500~MHz. \figlab{FW-stokes-200}}
\end{figure*}

The comparisons of \SName\ with CoREAS in the previous section have been done for the LOFAR frequency bandwidth of 30 -- 80~MHz. To test the suitability of the macroscopic approach in other bandwidths we have calculated the radio footprint for the shower used for \figref{FW-stokes-H} for two other bands.

In the 100 -- 200~MHz band one sees from \figref{FW-stokes-100} that the Cherenkov ring starts to emerge with a peak intensity in a broad ring at a distance of 50 -- 100~m. The intensity depends on azimuth angle due to the interference with charge-excess radiation. The results of \SName\ agree quite well with those of the CoREAS simulation. At distances exceeding 200~m the agreement worsens due to the fact that the CoREAS simulation creates a noisy signal where to noise, being independent of distance, at a certain point exceeds the signal. As a result the relation for the Stokes parameters, $Q^2 + U^2 + V^2 =I^2$ is no longer obeyed. This shows as a drop in $Q/I$ while $U/I$ and $V/I$ remail small. The \SName\ does not suffer from stochastic noise.

The Cherenkov ring is fully developed in the 200 -- 500~MHz band can be seen from \figref{FW-stokes-200}. The agreement between \SName\ and the CoREAS simulation is still very convincing, even though the differences in calculated intensities show a stronger systematic trend. At the higher frequency the problem with numerical noise is enhanced which is why the CoREAS simulation is no longer reliable beyond 150~m distance from the axis as well as at distances less than 50~m. At a distance near 100~m the circular polarization $V/I$ is negligible in the \SName\ calculation while CoREAS shows considerably larger values. We have not explored the source of this difference that seems to point to an underestimate of the difference in emission heights between charge excess and transverse current radiation in \SName.

\subsection{Footprint under thunderstorm conditions}

With increasing force on the electrons the power of the emitted pulse increases until a maximum is reached for a field of the order of 50~kV/m~\cite{Tri15}. The reason for this saturation is two fold; a) the transverse drift velocity is limited because the velocity of the particles cannot exceed $c$, as expressed in \eqref{Def-u}, and b) the pancake thickness increases with increasing field strength as shown in \figref{E-pancake}. It should be noted that these two effects are related since as the transverse velocity increases, the longitudinal component must decrease and thus the particles lag further behind the shower front. In \figref{E-FieldDep} the maximal peak power calculated using the semi-analytic approach is compared with the results of a microscopic calculation. The same two-layer field configuration is used as in Ref.~\cite{Tri15}, a top layer between 8 and 3~km with a net force strength $F$ and a lower layer from 3~km till the ground with a strength $0.3\times F$ in the opposite direction where the strength of $F$ is varied. For the shower $\Xmax=500$~g/cm$^2$ is taken. With the parameter $\upsilon_0$  set to the value given in \appref{Param} an excellent agreement is obtained.

\begin{figure}[ht]
\includegraphics[width=0.40\textwidth]{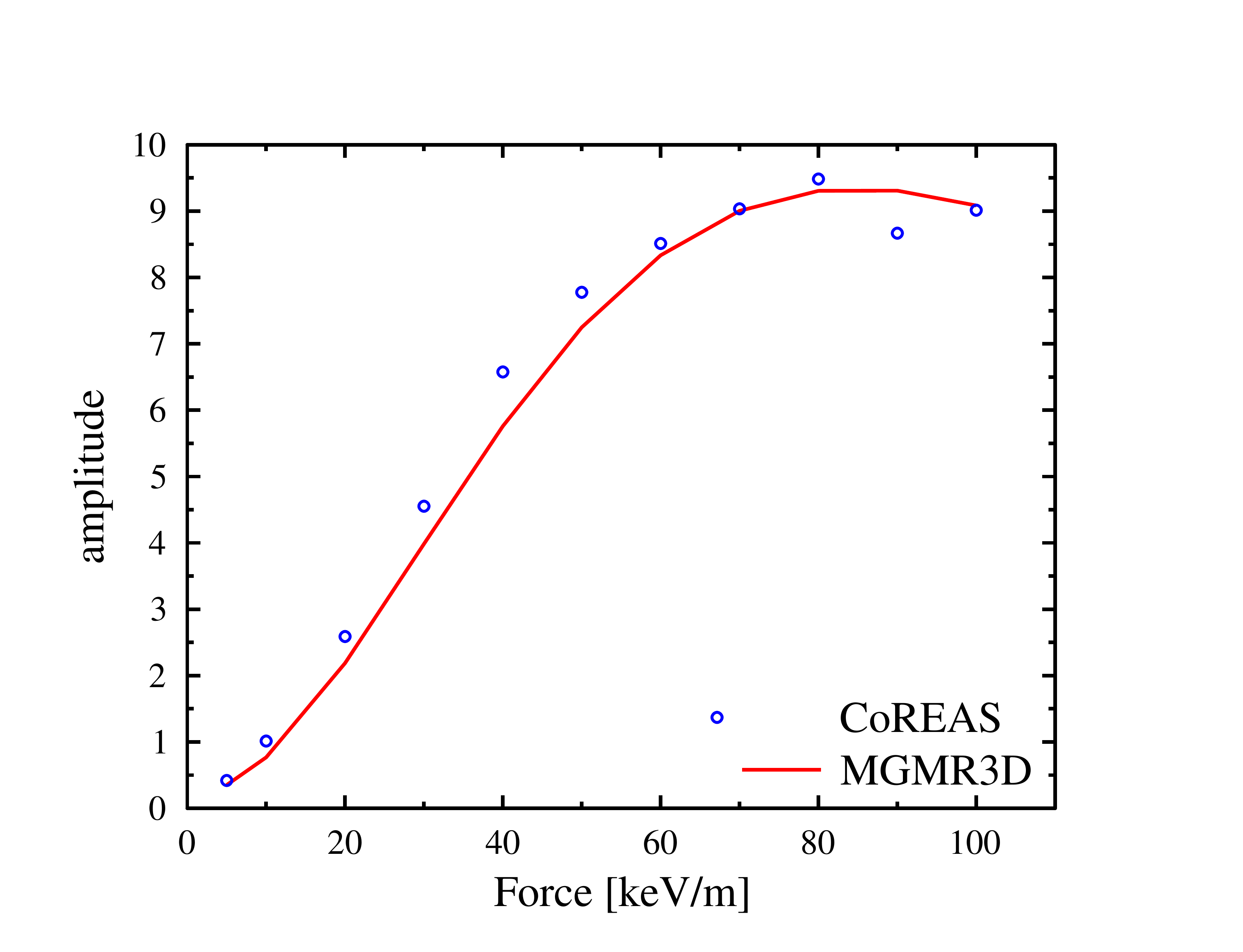}%
 \caption{Comparison of the dependence of the peak pulse power on the atmospheric electric field.  \figlab{E-FieldDep}}
\end{figure}

\begin{figure*}[t]
 \includegraphics[width=0.94\textwidth]{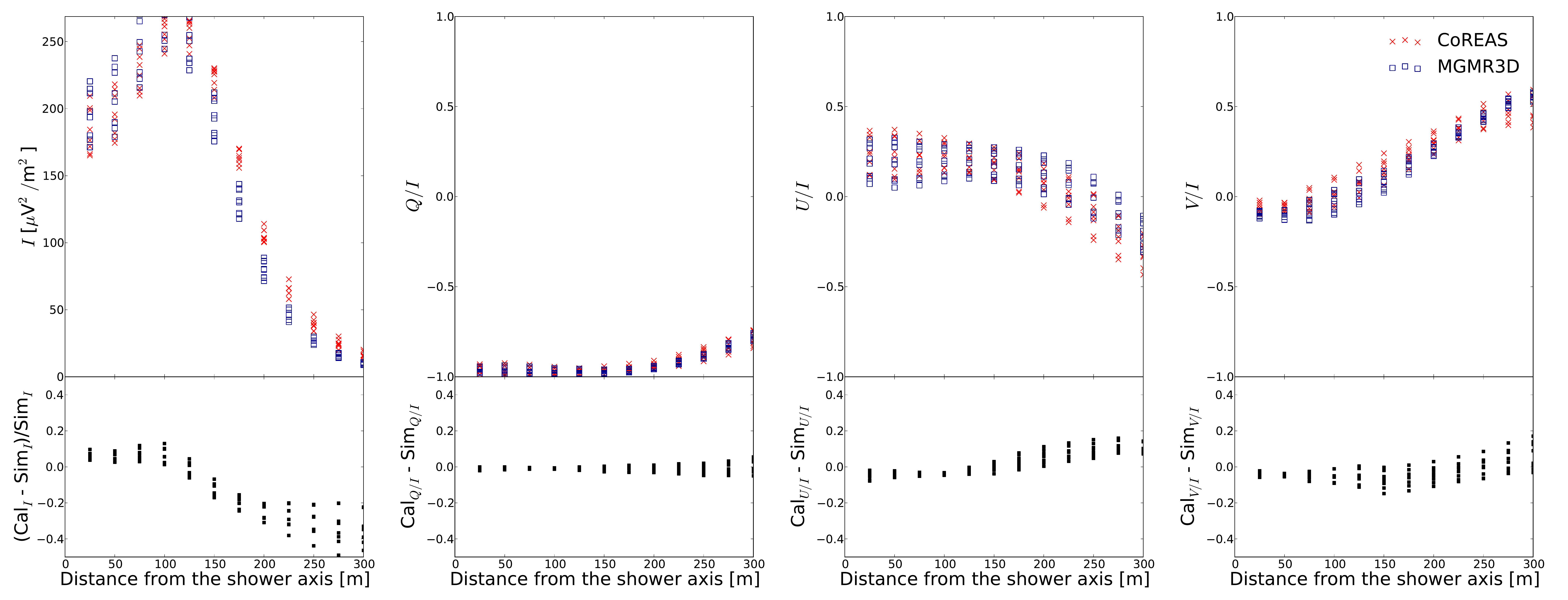}
 \caption{Comparison of stokes parameters with the the results of a microscopic calculation for a dual-layered atmospheric electric field, see \tabref{ThField} first column for the structure of the atmospheric electric field. \figlab{Th-stokes-2}}
\end{figure*}

\begin{figure*}[t]
 \includegraphics[width=0.94\textwidth]{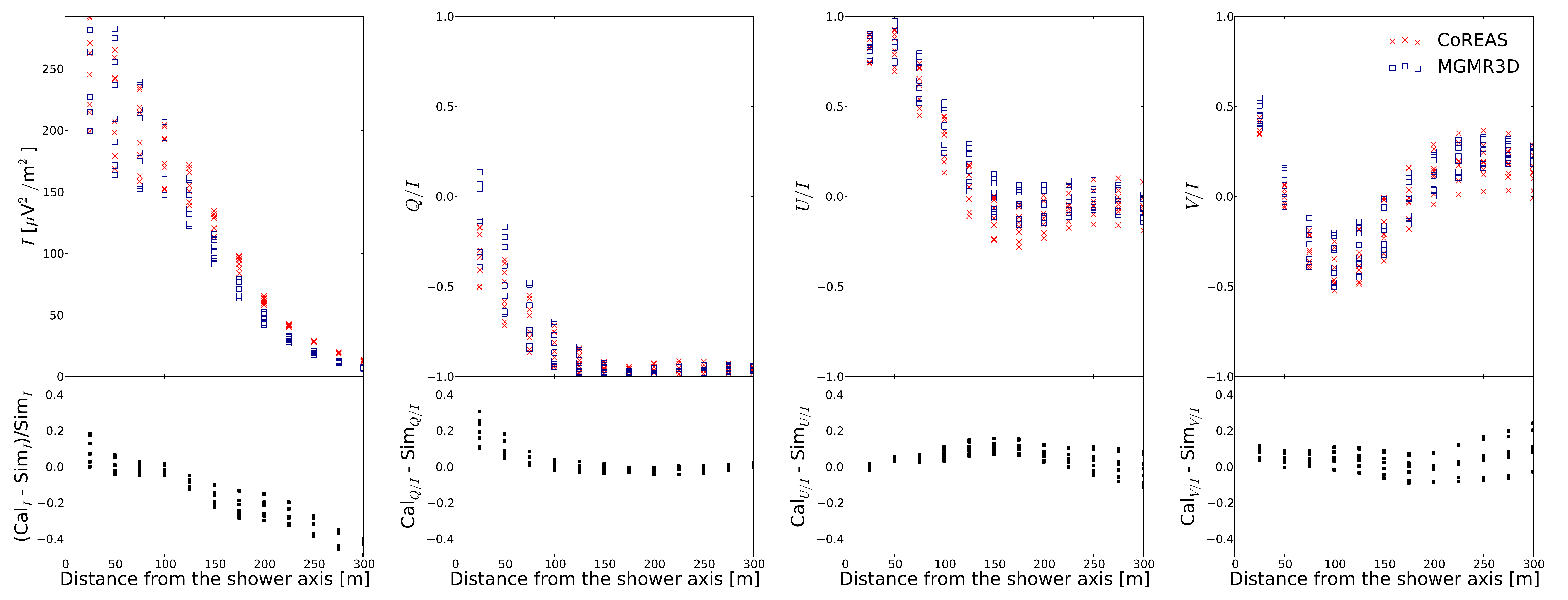}
 \caption{Same as \figref{Th-stokes-2} for a triple-layered atmospheric electric field, see \tabref{ThField} second column for the structure of the atmospheric electric field. \figlab{Th-stokes-3}}
\end{figure*}

\begin{table}[ht]
\caption{Two typical thunderstorm electric-field configurations used in the examples. Listed are the heights at which the force changes in magnitude and direction. $\alpha$ denotes the angle of the net force with the Lorentz force due to the magnetic field of Earth.  \tablab{ThField}}
\begin{tabular}{|l l|c|c|}
\hline
\multicolumn{2}{|c|}{Configuration} & \figref{Th-stokes-2} &\figref{Th-stokes-3}\\
\hline
Layer 1& $h_1$ [km] & 8.0 & 8.0\\
        & $F_1$ [keV/m] & 50 & 50\\
        & $\alpha_1$    & 90$^\circ$ & 90$^\circ$\\
Layer 2 &$h_2$ [km] & 3.0 & 5.0\\
        & $F_2$ [keV/m] & 15 & 15\\
        & $\alpha_2$    & 270$^\circ$ & 90$^\circ$\\
Layer 3 &$h_3$ [km] & -- & 3.0\\
        & $F_3$ [keV/m] & -- & 15\\
        & $\alpha_3$    & -- & 0$^\circ$\\
\hline
\multicolumn{2}{|c|}{$\Xmax$} & 510~g/cm$^2$ & 660~g/cm$^2$ \\
\hline
\end{tabular}
\end{table}

In \figref{Th-stokes-2} and \figref{Th-stokes-3} the footprint is compared to the results of a CoREAS simulation for two different realistic atmospheric electric field configurations as can be expected during thunderstorms as given in \tabref{ThField}. In \figref{Th-stokes-2} the electric fields in the two layers have opposite directions and the emission of the top and bottom layer interfere destructively near the core which becomes less efficient with increasing distance due to the finite refractivity of air. As a result a ring-like structure is seen in the intensity. Beyond the ring the intensity falls off a bit faster in the MGMR3D calculation as compared to CoREAS. We have investigated if this could be corrected for by changing the heights at which the electric fields change, however this did not yield satisfactory results.
The emission is mainly linearly polarized along the direction of the net force, which for this case is perpendicular to the $\vB$ axis. This results in Stokes $Q/I\simeq -1$ and $U/I\simeq 0$. Since the transverse current is much larger than for a typical fair weather shower due to the strong electric field, the relative contribution of charge excess radiation is very small which is reflected in a much smaller spread in the values of $U$. The circular polarization of the pulse near the core is small, $V/I \simeq 0$ because the electric fields in the two layers lie in the same plane.

The observed structure is very different from the field configuration used in \figref{Th-stokes-3}.
Near the core the signal from the lower layer arrives before the signal emitted from the higher layers since the shower proceeds with the speed of light while the radio signal propagates at a reduced speed due to the finite refractivity of air. This results in a large circular polarization near the core, $V/I \simeq 0.5$.
At a distance of 100~m from the core, due to different traveling distances, the situation is reversed and the signal from the upper layers arrives before that of the bottom layer resulting in a reversed circular polarization, $V/I \simeq -0.5$. Because of the changing relative importance of the different electric-field layers an intriguing dependence of the linear polarization with distance is observed. Near the core the net polarization is oriented at an angle of 45$^\circ$ with respect to the $\vB$ axis, giving rise to $Q/I\simeq 0$ and $U/I\simeq +1$. At larger distances the radiation from the top layer dominates resulting in a linear polarization normal to $\vB$ ($Q/I\simeq -1.0$ and $U/I\simeq 0.$) and vanishing circular polarization, $V/I\simeq 0$.
The intensity shows a strong peak near the core, like is seen for fair weather events, since for the field configuration of \figref{Th-stokes-3} there is no destructive interference.
It is seen that for these rather complicated atmospheric-field configurations the results of semi-analytic calculation lie very close to those of the microscopic calculation.

\section{Conclusions}

With a relatively simple parametrization of the structure of the leading plasma cloud in an EAS we are able to recover the main characteristics of the radio emission in an analytic calculation with \SName.  The basic structure of the emitted pulse and the complete polarization footprint follow the result of a more complete microscopic calculation. This holds over a wide range of frequencies and some non-obvious structure of the height dependence of the induced currents in this cloud.

On the one hand this gives us a better insight into the physics that is important for understanding radio emission. One finding is that the typical pulse shape, a large peak followed by a long tail of opposite polarity, appears due to the radial dependence of the pancake thickness. In the extreme that the pancake thickness has no radial dependence a bi-polar pulse is obtained with comparable strengths for the two polarities. Another interesting point is that one might have expected that the induced transverse currents in the plasma would result in a net dipole charge distribution in the plasma cloud, moving with the speed of light. We have not seen any evidence for such a contribution to the radio pulse. A finite contribution of such a term in the \SName\ calculation will give rise to a worse agreement with the results of the microscopic calculations.

We are able to rather accurately predict the structure of the radio footprint for rather complicated structures for the height dependence of the induced current using relatively few parameters for the structure of the charge cloud that are kept constant. We see this as a reflection of shower universality. Because of this it is feasible to use \SName\ in an optimization code to extract the transverse current structure by fitting the results of an \SName\ calculation to data. To this end the \SName\ code has been implemented in a Levenberg-Marquardt minimization procedure, that is based on a steepest descent method, to extract the current distribution in the atmosphere during thunderstorms.

\appendix
\section{Calculation of the radiation field}

The radiated electric field is derived from the vector potential using
\beq
\vec{E}_i(t,\vec{x})=-{\partial\over \partial \vec{x}_i} A^0 -{\partial\over \partial t}\vec{A}_i \;.
\eeq
A difficulty in evaluating the radiation fields lies in the fact that the retarded distance ${\cal D}=0$ lies in the integration regime. To regularize this one can efficiently use partial integration techniques as shown in Ref.~\cite{Wer08}.

To evaluate the integration over $z'$ in \eqref{L-W} we follow the same approach as used in Ref.~\cite{Wer12} and replace the integral in the $z$-direction by an integral over $\lambda=h_c-h$ where at the critical height, $h_c$, ${\cal D}=0$. This substitution allows for an easier calculation of derivatives of the vector potential that are necessary to calculate the electric field. When going from the expression for the vector-potential to the electric fields the derivatives on the $\lambda$ integration limits vanish. Note that the retarded time in \eqref{L-W} depends on $\lambda$ as well as the other integration variables, as shown in Ref.~\cite{Wer08,Wer12}

In calculating the radiation fields we distinguish the charge excess and the transverse current contributions, denoted by the superscripts CX and TC, respectively. The full radiation field is the vectorial sum of the two. The current distributions in the plasma cloud are defined following \eqref{DefCloud}.

\subsection{Charge excess}\seclab{CE}

The charge excess in the shower front is given by $J_Q(z)$, as defined by \eqref{Def-IQ}, propagating with the speed of light in the $-z$-direction thus contributing to the zero and the $z$ component of the vector potential, \eqref{L-W}. Because of the axial symmetry the electric fields are given by
\bea
E_r&=&-\partial\, A^0/\partial r_o \nonumber \\
E_z&=& -{\partial\over \partial z_o} A^0 -{\partial\over \partial t_o}A^z \;,
\eea
where $r_o=d$ is the distance from the observer to the shower axis.

Substituting the expression for the vector potential, integrating over the spatial extent of the charge cloud the radially polarized radiation field can be written as
\bea
E_r^{CX} &=& -{\partial A^0\over \partial r_o} =
  -\!\int\! dx_s\,dy_s\,{\partial w(r_s)/r_s\over \partial r_s}\, I^{CX}  \nonumber \\
  &-& \!\int\! dx_s\,dy_s\, w(r_s) \!\int\! dh \,{\partial f(h,r_s)\over \partial r_s}\, \left.{J_Q \over {\cal D}}\right|_{h} \;, \eqlab{E_cx}\\
I^{CX} &=&\int\! dh \, f(h,r) \left.{J_Q \over {\cal D}}\right|_{h} \;,\\
{\partial f(h,r_s)\over \partial r_s} &=& -(1+\alpha-2 h/\lambda) \, {f((h,r_s)\over \lambda} {d\lambda\over d r_s}\;,\eqlab{I_cx}
\eea
where $r_s^2=(x_s^2+y_s^2)$, the retarded distance ${\cal D}$ is given by \eqref{denom}, $J_Q$ is the net charge as given by \eqref{Def-IQ}, and the defining \eqref{DefCloud} is used to introduce the profile functions $f(h,r_s)$ (\eqref{Def-f}) and $w(r_s)$ (\eqref{Def-w}). The integral is separated into an integration over $h$, a 'ray', a line parallel to the shower axis, of the full shower where special care should be devoted to the point where ${\cal D}=0$, followed by one over the radial extent. In the numerical calculation the results for separate rays, $I^{CX}(t)$, are stored for a grid of $r_s$ and $d$ values, where the latter is the distance of the observer to the single ray.

\subsection{Transverse current}\seclab{TC}

Along a similar line of reasoning as for the charge-excess radiation, the transverse current radiation field is written as
\bea
E_x^{TC} &=& -{\partial A^x\over \partial t^o} =\int dx_s\,dy_s {w(r_s)\over r_s} I_x^{TC} \;, \eqlab{E_tc} \\
I_x^{TC} &=& \int_0^{h_c} dh \, f'(h) \left.{J^x \over {\cal D}}\right|_{h} 
 - \int_0^{h_c} dh\, f(h) \left.{J^{\prime x} \over {\cal D}}\right|_{h} \;.\eqlab{I_tc}
\eea
with a similar expression for the component polarized in the $y$ direction. Similar to the notation introduced before we have $J^{\prime x}=dJ^x /dt_r$ and $f'(h)=df/dh$. The results for rays, $I^{TC}(t)$, are calculated once on a grid of $r_s$ and $d$ values to be used subsequently in the calculation of the complete footprint of the electric field.

\section{Parameterizations}\applab{Param}

\begin{table}[!ht]
\caption{Determined parameter values\tablab{Params}}
\begin{tabular}{|r l c|r l|l|}
\hline
$X_0$ & \eqref{Def-Nc} &   & 36.7 & [g/cm$^2$] \Prog{X-0}\\          
$\gamma$ & \eqref{Def-Nc} &   & 90.0 & [g/cm$^2$] \Prog{ lamx}\\       
$M_0$ & \eqref{Def-w} & \figref{w} & 27.0 & [m] \Prog{MoliereRadius}\\                    
$\Lambda_0$  & \eqref{Def-lamr} & \figref{jrh} &  0.05 &  [m] \Prog{ lam-tc}\\   
$\Lambda_1$  & \eqref{Def-lamr} & \figref{jrh} &  7. &  [m] \Prog{lam-100}\\     
$F_\beta$ & \eqref{Def-sigma} & \figref{Jxz}-a &  300 & [keV/m] \Prog{F-over-beta} \\      
$a_t$ & \eqref{Def-sigma} & \figref{Jxz}-a &  2. &   \Prog{?} \\
$a_c$ & \eqref{Def-rhc} & \figref{Jxz}-b &  0.5 &  \Prog{?} \\
$\upsilon_0$ & \eqref{Def-u} & \figref{E-FieldDep} &  0.2 &  \Prog{ (U0=60)/F-over-beta} \\              
$a_E$  & \eqref{Def-alphaE} & \figref{E-pancake} & 0.41 & \Prog{PancakeIncField}\\                  
\hline
\end{tabular}
\end{table}

The values of the parameters in the parametrization of the plasma cloud, co-moving with the shower, are given in \tabref{Params}, including the defining equation. All parameters have been determined by fitting the results of a Monte Carlo simulation and the table specifies the figure where the fit is shown. The only exception are the parameters for the longitudinal shower profile, $X_0$ and $\gamma$.

\section{Programming details}

In the numerical implementation we have exploited the axial symmetry of the current densities as much as possible in order to optimize the running time for the calculation of a single footprint. The integration over the plasma cloud, \eqref{E_cx} and \eqref{E_tc}, is performed in two steps. In the first step an integration is performed along a single `ray', i.e. \eqref{I_cx} and \eqref{I_tc}, at a fixed distance from the shower axis, $r_s$, and on a grid of distances from this ray to the observer, $r_{so}$. Since this is done separately for the charge excess and the transverse current contributions the dependence of the components on the azimuth angles are thus simple sine or cosine functions which will be taken into account at the next stage. These ray-integrals are stored on a three-fold grid on $r_s$, $r_{so}$, and time values. The final integral for an observer at a distance, $r_o$, from the core of the shower runs over $r_s$ and azimuth angles $\phi_s$ where the distance $r_{so}$ is calculated using straight-forward geometry. Thus the radial component (due to charge excess) and the two linearly polarized components (due to transverse currents in the $\vB$ and the $\vvB$ directions) of the radiation field are calculated separately on a grid of observer distances, $r_o$. The calculation of the radiation field for a single antenna can be obtained through an interpolation on the $r_o$-grid and vectorially adding the different contributions.
Due to this procedure all angle integrals are almost trivial. The integral for a single ray is relatively expensive since retarded distances have to be calculated with care.

At many levels in the calculation an interpolation is necessary to obtain a time-dependent signal for a distance between two grid points. To be able to perform an efficient interpolation it was realized that pulses in subsequent grid points generally have a very similar time structure, however, with a relative time shift. The interpolated pulse can thus be approximated most accurately by averaging the time structures, correcting for the relative time shift and subsequently applying an interpolated time shift. This allowed for using a relatively sparse radial grid.

As an alternative to this procedure we initially used an interpolation of the fourier components of the pulses. For cases where the relative time shifts amounted to half the spacing on the time grid this procedure introduced unrealistically strong high-frequency components for which reason we abolished it.

The full calculation of a radio footprint in MGMR3D takes of the order of 5 CPU seconds, independent of energy and almost independent of the number of antennas, to be compared with approximately a CPU day for CoREAS for the same footprint at E=$10^{16}$~eV and 160 antennas. The time for a CoREAS calculation, for unchanged thinning, increases linearly with energy. The MGMR3D code is available upon request from the authors.

\end{document}

%% file: Preambule-Refs.tex
\def\VYPx#1#2#3{{\bf #1}, #3 (#2)}  
\newcommand{\VYP}[4][]{%
  \ifx\\#1\\
    \VYPx{#2}{#3}{#4}
  \else
    \href{http://dx.doi.org/#1}{\VYPx{#2}{#3}{#4}}
  \fi
}

\def\etal{ {\it et al.}}
\def\Ttl#1{``{\it #1}'',}

\def\PRD#1#2#3{\href{http://link.aps.org/doi/10.1103/PhysRevD.#1.#3}{Phys.~Rev.~D~\VYP{#1}{#2}{#3}}}

\def\PRL#1#2#3{\href{http://link.aps.org/doi/10.1103/PhysRevLett.#1.#3}{Phys.~Rev.~Lett~\VYP{#1}{#2}{#3}}}

\newcommand{\AeA}[4][]{Astron.~\& Astrophys.\ \VYP[#1]{#2}{#3}{#4}}

\newcommand{\ApP}[4][]{Astropart.~Phys.\ \VYP[#1]{#2}{#3}{#4}}

\newcommand{\JCAP}[4][]{J.~Cosm.~and~Astrop.~Phys.\ \VYP[#1]{#2}{#3}{#4}}

\def\arXiv#1{\href{http://arxiv.org/abs/arXiv:#1}{arXiv:{#1}}}
\def\nucl-th#1{\href{http://arxiv.org/abs/nucl-th/#1}{nucl-th/{#1}}}